# A High Precision Signals Readout System for Micromegas Detector Based on the VMM

Shuang. Zhou, Shengquan. Liu, Peng. Miao, Xinxin. Wang, Feng. Li, Ge. Jin, Zhilei. Zhang, and Tianru. Geng

***Abstract*–Micromegas detector is a gas detector with parallel plate structure, and it consists of a conversion gap in which radiations liberate ionization electrons and a thin amplification gap. The signal of the micromegas detector consists of two parts: the electron peak and the ion tail. The electronic signal just keeps few nanoseconds, which is used for precise time measurement. The ion signal carries most of the signal energy, and it is used to reconstruct particle energy. The micromegas detector has the advantages of high counting rate, high gain, good time resolution and position resolution, excellent radiation-hardened performance, large sensitive area and readout convenience. This paper introduces a signals readout electronics system for micromegas detector based on VMM chips. The VMM is the front end ASIC to be used in the front end electronics readout system of both the micromegas and sTGC detectors of the New Small Wheels Upgrade project. The VMM is designed by Brookhaven National Laboratory, and each chip is composed of 64 linear front-end channels. Each channel integrates a low-noise charge amplifier (CA) with adaptive feedback, test capacitor, and adjustable polarity (to process either positive or negative charge). Based on those characteristics, the VMM is applicable for the signals readout system of multichannel detectors. The readout system consists of three parts: the front-end board, the data acquisition board, and the host computer with the control software.**

## I. Introduction

Compared with other types of detectors, high resolution of time and position is one of prominent characters of the micromegas detector. Micromegas detector is a parallel plate structure gas detector, and it consist of a drift electrode, a micromesh and signal strips. When a cosmic ray or a charged particle goes through the detector, a larger number of electron-ion pairs are generated in the detector's conversion gap. Because of the drift electric field, electrons move to the amplification gap where the avalanche breakdown occurs. Because the drift electrode and the micromesh are connected to different negative voltage, and signal strips are connected to the ground, therefore, the charge polarity of the strips is negative.

There are three important parameters to value the performance of micromegas detector: energy resolution, time resolution, and position resolution. Both of the detector and the readout electronics system have obvious impacts on resolutions. High energy resolution requires the readout system can sample signals with high precision; high time resolution requires the readout system have a short dead time; high position resolution means high density of signal strips, so it requires readout system can process multi-channel signals and a huge mess of signal data. The VMM is an appropriate front-end ASIC to realize the micromegas readout system. It has adjustable signal polarity and 64 linear front-end channels. For sampling signals with high precision, the VMM integrate lots of high precision ADCs. This paper introduces a high precision readout system for micromegas detector, and the whole system has three main parts: front-end electronics based on the VMM, data acquisition system and the control software.

## II. VMM Chip

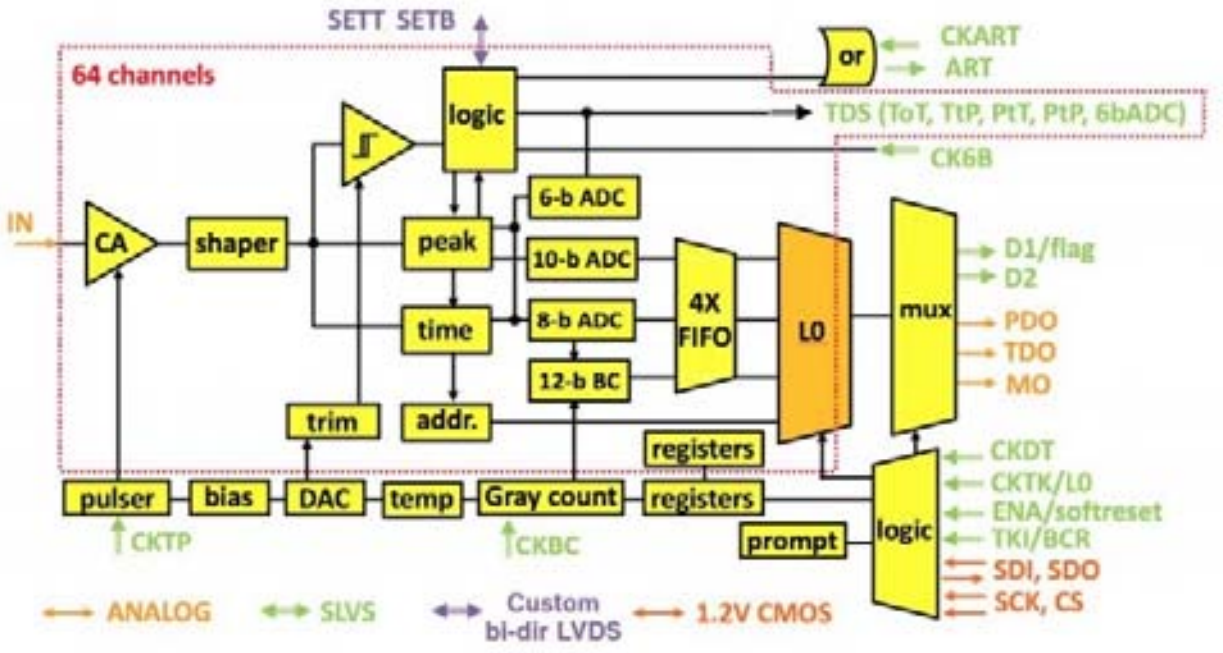

Fig. 1. The architecture of the VMM3 chip.

The VMM3 chip is the third generation of the VMM. The architecture of the VMM3 chip is illustrated in Fig. 3. The VMM3 chip can operate in three modes. In our design, continuous mode is applied. The VMM3 is composed of 64 front-end channels. Each channel integrates a low noise charge amplifier, a shaper circuit with a delayed dissipative feedback (DDF), test capacitor, and adjustable polarity. Next to the shapers are the sub-hysteresis discriminators, the peak detector, and the time detector. The discriminators discriminate the signals exceeding the threshold. The peak detector measures the peak amplitude and stores it in an analog memory. The time detector measures the timing using a time-to-amplitude converter (TAC). The VMM3 chip also features a baseline stabilizer circuit and the gain is configurable. The global test pulser DAC provide the ability to send test pulse signals to each individual readout channel for calibration.

After the whole measure process of VMM3, a total of 38 bits data are generated for each event. The 38 bits are consist of one bit flag which can be used to open a readout window, one bit threshold crossing indicator, six bits channel address, ten bits peak amplitude (PDO) and twenty bits timing information. The peak amplitude information is the measure result of a 10-bit ADC, and the timing information is the measure result of a time-to-amplitude converter (TAC) based on a 6-bit ADC. Besides peak amplitude and timing information, VMM3 also offers an analog output monitoring output (MO), which is used to monitor lots of analog signals of the VMM3, such as channel analog signal, threshold voltage, and temperature sensor result.

## III. READOUT SYSTEM

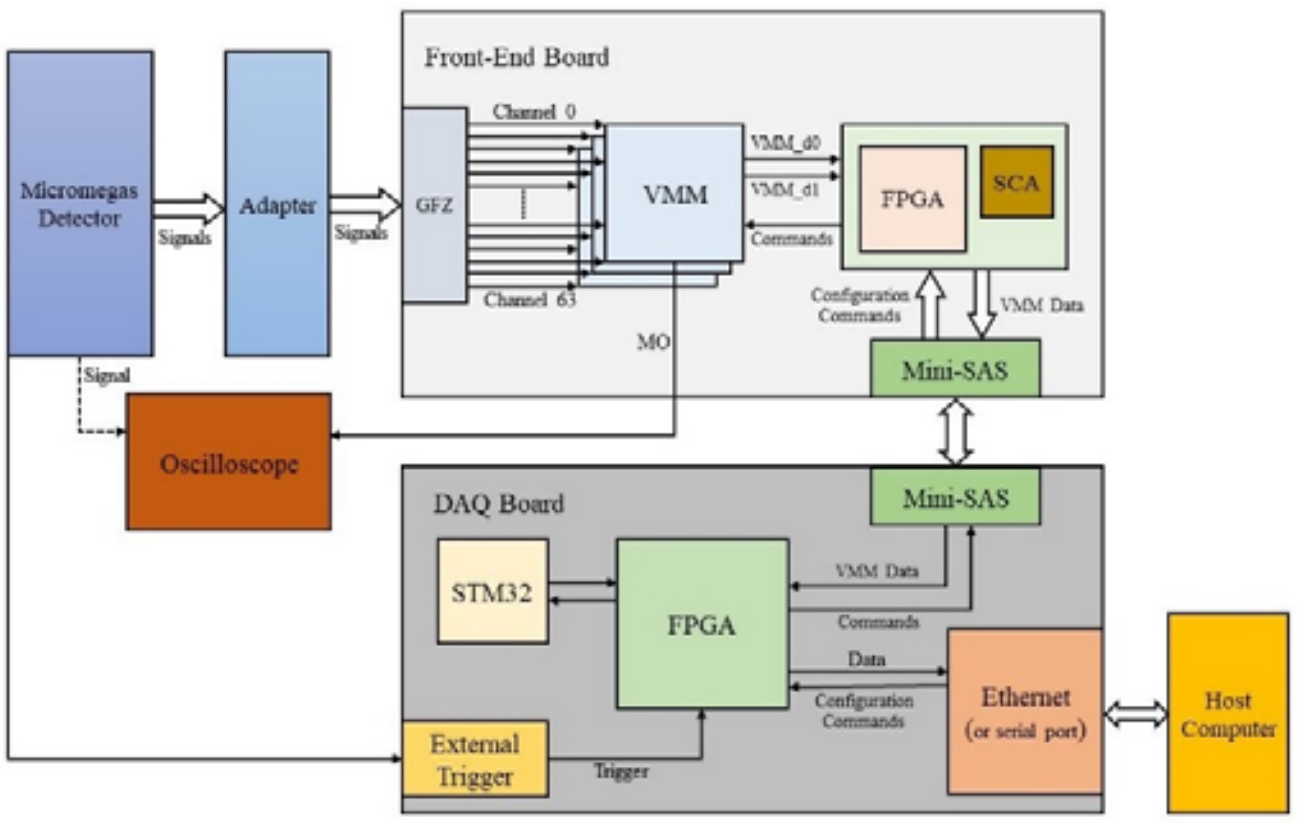

Fig.2. Schematic block diagram of the readout system

The simplified block diagram of the readout electronics system for micromegas detector based on the VMM is illustrated in Fig.2. When a cosmic ray or a charged particle goes through the detector, the detector output corresponding signals to the front-end board, and then these signals will be measured by VMM3 chips on the front-end board. The FPGA on the front-end board buffer the data from VMM3 chips, and send the data to the data acquisition board (DAQ board). The DAQ board wrap the VMM3 data, and send the data to host computer through Ethernet. This is a whole data readout process of the readout system.

Before the readout process, users should configure the VMM3 chips at first. Users set needed configurations at the control software on the host computer, and send configuration commands to control the working mode of VMM3 chips. The DAQ board is in charge of the commands communication between FEB and host computer, because the front-end board can't connect with host computer directly.

### A. Front-End electronics Board

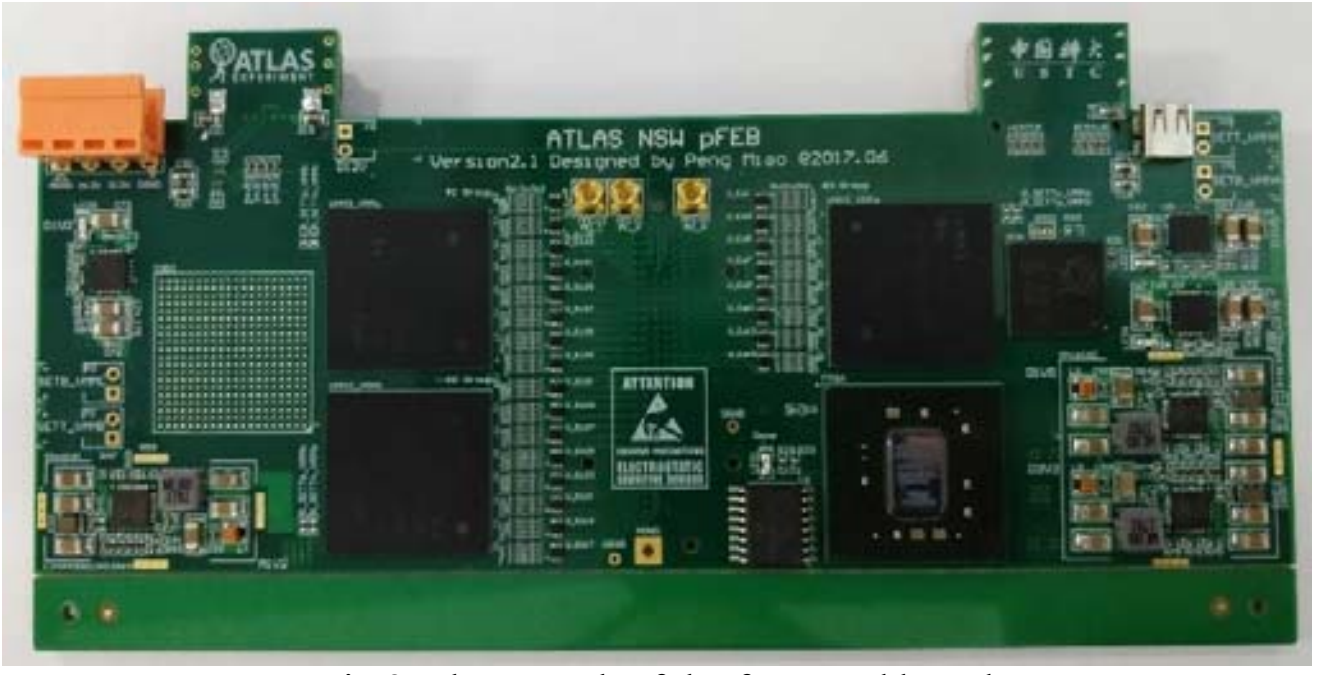

Fig.3. Photograph of the front-end board

The product photograph of the front-end board is shown in Fig. 3. The front-end board mainly consists of a GFZ connector used as detector signals input ports, three VMM3 chips, a FPGA, a slow control adapter (SCA), and two mini-SAS connectors. VMM3 chips send out measure results of the signals through VMM_d0 and VMM_d1 two data lines to the FPGA. The FPGA on the front-end board analysis and pack the data, through the mini-SAS cable, signals data are sent to the DAQ board. The SCA is mainly used for configuring VMM3 chips. Besides the digital signal data, there is also an access to observe each channel's analog signal through the MO port.

There are three VMM3 chips (192 channels) on this type front-end board, for different detector requirements, changing the number of VMM3 chips to control the number of front-end electronics channels.

### B. Data Acquisition Board

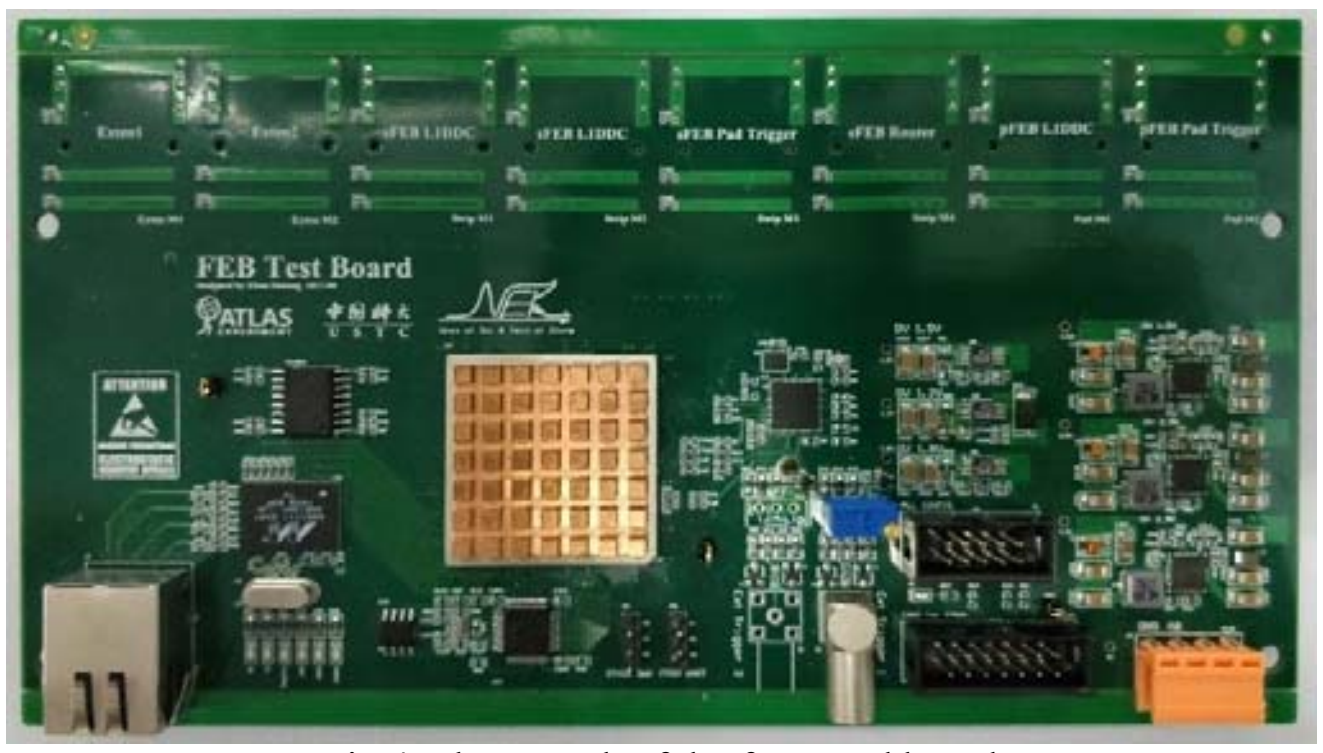

Fig.4. Photograph of the front-end board

The product photograph of the data acquisition board (DAQ board) is shown in Fig. 4. The DAQ board mainly consists of mini-SAS connectors, a FPAG, a STM32 chip, an Ethernet module, and an external trigger module. The FPGA on the DAQ board receives VMM data from mini-SAS connectors, and then analysis and wrap the data to meet Ethernet transmission requirements. The wrapped data are sent to the host computer for analyzing or storing.

During the configuring process, configuration commands from host computer are decoded on the DAQ board, with the cooperative work of the STM32 chip and the FPGA, and then the FPGA send decoded commands to the front-end board to configure VMM3 chips.

Besides, there is an external trigger module on the DAQ board to control the data acquisition process. The external trigger comes from micromegas detector, and it is a start signal for the DAQ board to read out signal data.

### C. Control Software

As shown in Fig. 5, users can conveniently set VMM3 configurations and acquire data on the control software GUI (Graphical User Interface). During the experiment, user just need to control the GUI, don't have to adjust the physical readout system.

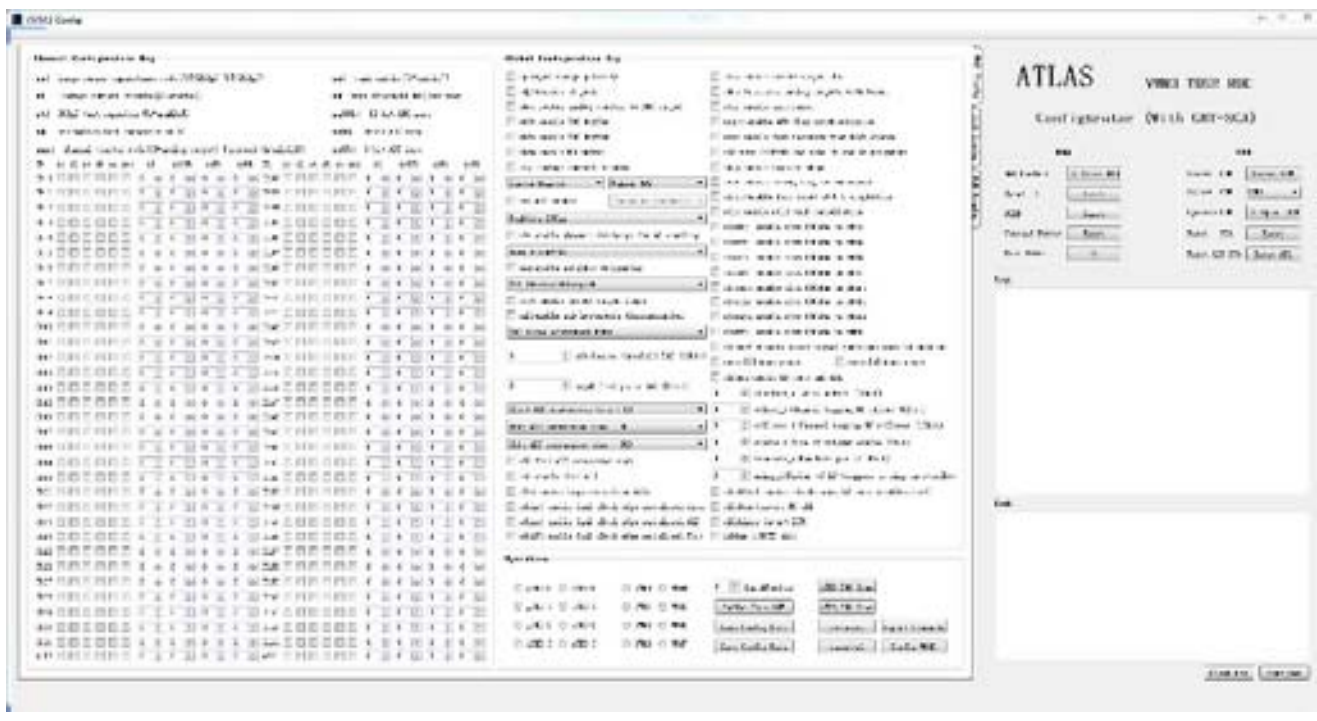

Fig.5. GUI of the control software

## IV. Readout System Performance

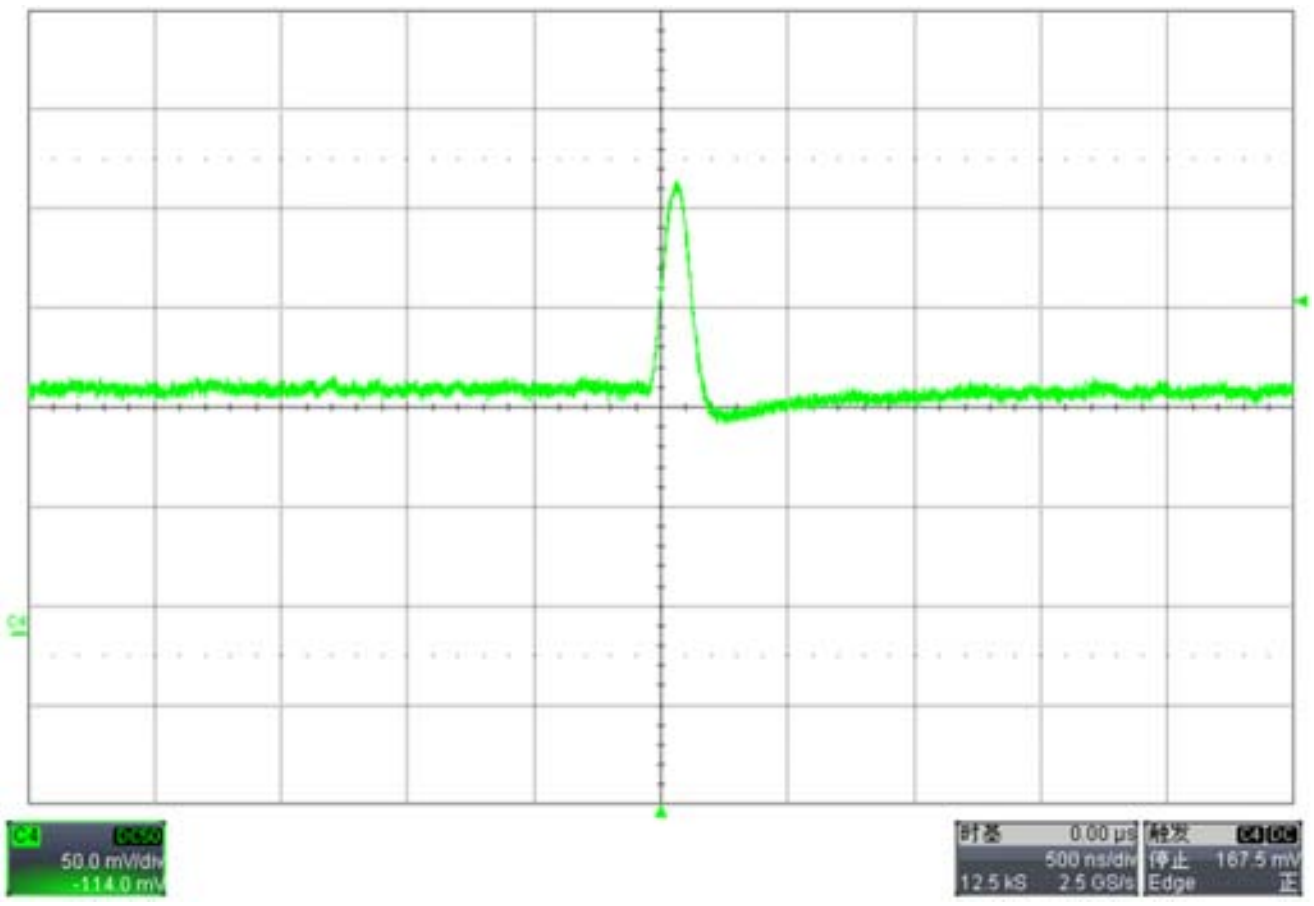

Fig.6. Analog signal output of a VMM3 channel

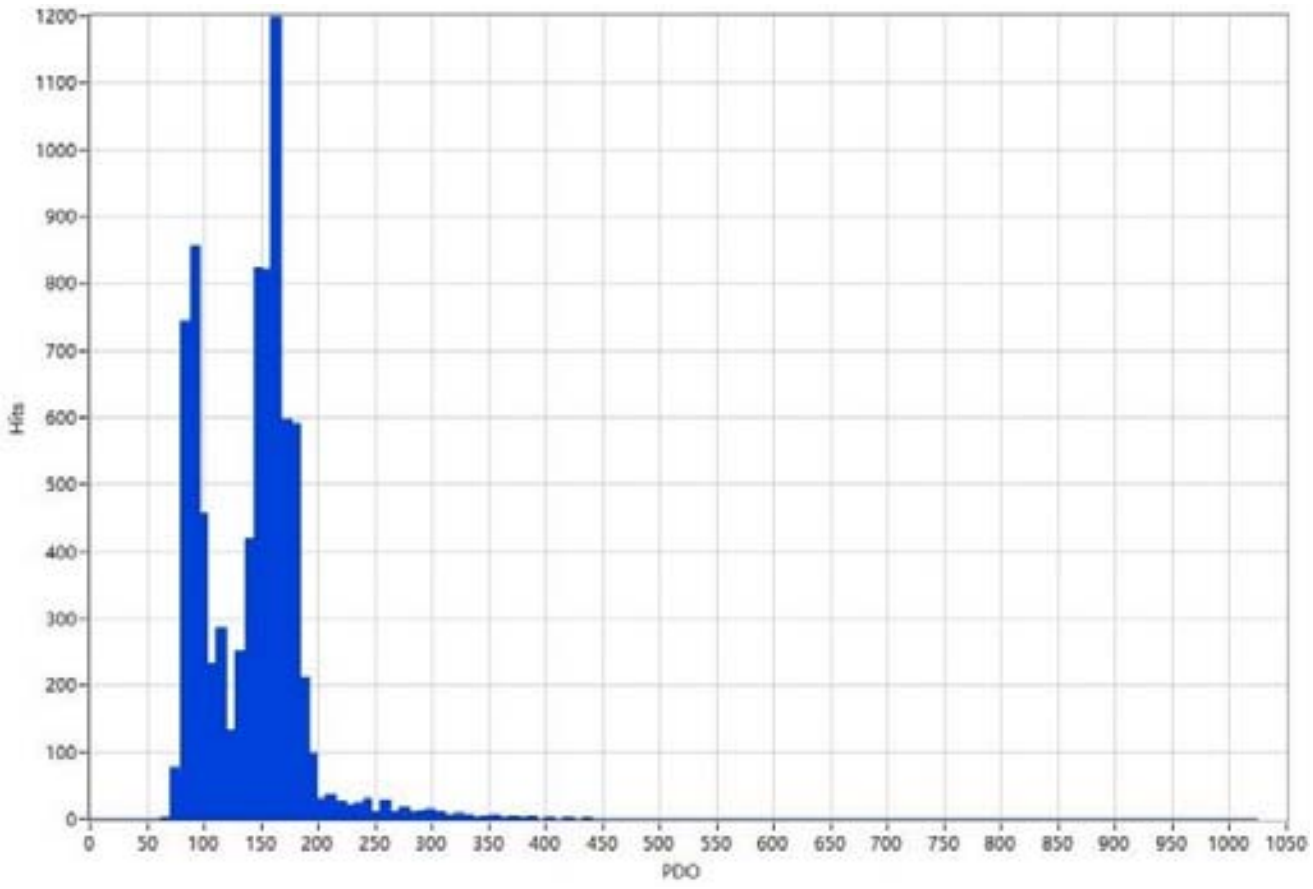

Fig.7. Energy spectrum of $Fe^{55}$ X-ray

As shown in Fig.6, a signal from micromegas detector cause analog outputs of on VMM3 channels. The output signals have good SNR (signal to noise ratio) and good shaping result. Fig. 7 shows an energy spectrum of $Fe^{55}$ X-ray which is the statistics result of the readout data from this readout system. Analyzing this energy spectrum, the energy resolution is 20.6%. The energy resolution can reach a lower degree by adding test time and adjusting VMM3 configurations.

For VMM3 chips, each channel has a minimum 200 ns dead time after a peak is found, which can totally meet the requirement of micromegas detector to realize the high time resolution. For VMM3 chip integrates 64 front-end channels, it's quite convenient to realize high channel density readout system, which can ensure the high position resolution of micromegas detector.

## V. Conclusion

In this paper, a high precision readout system for micromegas detector has been described in detail. The whole readout system include three important parts: front-end electronics, data acquisition system and the control software. The core of the front-end electronics is the VMM3, and the data transmission between the host computer and the data acquisition system uses Gigabit Ethernet. In view of the applicability of VMM3 chips and this readout system, this high precision readout system can also be used in other kinds of detectors.